\documentclass{elsart}

\usepackage{graphicx}
\usepackage{amsmath}
\usepackage{amssymb}
\usepackage{amsfonts}
\usepackage{latexsym}
\usepackage{theorem}
\usepackage{bbm}
\usepackage{bm}

\newtheorem{define}{Definition}
\newtheorem{theo}{Theorem}
\newtheorem{lemma}[theo]{Lemma}
\newtheorem{propo}[theo]{Proposition}

\newcommand{\ket}[1]{\ensuremath{|#1 \rangle}}
\newcommand{\bra}[1]{\ensuremath{\langle #1|}}
\newcommand{\bk}[1]{\ensuremath{\langle #1 | #1 \rangle}}

\newcommand{\1}{\protect\ensuremath{\mathbbm{1}}}

\newcommand{\abs}[1]{\ensuremath{| #1 |}}
\newcommand{\cb}[1]{\ensuremath{\| #1 \|_{cb}}}
\newcommand{\norm}[1]{\ensuremath{\| #1 \|_{}}}

\newcommand{\C}{\ensuremath{\mathbb{C}}}

\newcommand{\N}{\ensuremath{\mathbb{N}}}

\newcommand{\tr}[1]{\ensuremath{{\rm tr}(#1)}}
\newcommand{\trace}{\ensuremath{{\rm tr \,}}}

\newcommand{\id}{\ensuremath{{\rm id} \, }}

\newcommand{\re}[1]{\ensuremath{{\rm Re}(#1)}}

\newcommand{\hh}{\ensuremath{\mathcal{H}}}
\newcommand{\bh}{\ensuremath{\mathcal{B(H)}}}

\newcommand{\bc}[1]{\ensuremath{\mathcal{B}(\mathbb{C}^{#1})}}

\newcommand{\bhstar}{\ensuremath{\mathcal{B_{*}(H)}}}
\newcommand{\sbhstar}{\ensuremath{\mathcal{B}_{*,1}^{+}\mathcal{(H)}}}

\newcommand{\kk}{\ensuremath{\mathcal{K}}}
\newcommand{\bkk}{\ensuremath{\mathcal{B(K)}}}

\newcommand{\A}{\ensuremath{\mathcal{A}}}
\newcommand{\B}{\ensuremath{\mathcal{B}}}
\newcommand{\D}{\ensuremath{\mathcal{D}}}

\newcommand{\M}{\ensuremath{\mathcal{M}}}
\newcommand{\n}{\ensuremath{\mathcal{N}}}
\newcommand{\calS}{\ensuremath{\mathcal{S}}}

\begin{document}

\begin{frontmatter}



\title{A Continuity Theorem\\ for Stinespring's Dilation}


\author[tubs,pavia]{Dennis Kretschmann\thanksref{dennis}},
\thanks[dennis]{E-mail: {\tt d.kretschmann@tu-bs.de}}
\author[tubs,isi]{Dirk Schlingemann\thanksref{dirk}}, and
\thanks[dirk]{E-mail: {\tt d.schlingemann@tu-bs.de}}
\author[tubs]{Reinhard F. Werner\thanksref{reinhard}}
\thanks[reinhard]{E-mail: {\tt r.werner@tu-bs.de}}
\address[tubs]{Institut f\"ur Mathematische Physik, Technische Universit\"at Braunschweig, Mendelssohnstr.~3,
38106 Braunschweig, Germany}
\address[pavia]{Quantum Information Theory
Group, Dipartimento di Fisica A.~Volta, Universit\`{a} di Pavia,
via Bassi 6, 27100 Pavia, Italy}
\address[isi]{ISI Foundation, Quantum Information Theory Unit,\\
 Viale S. Severo 65, 10133 Torino, Italy}


\begin{abstract}
We show a continuity theorem for Stinespring's dilation: two
completely positive maps between arbitrary C$^*$-algebras are
close in cb-norm iff we can find corresponding dilations that are
close in operator norm. The proof establishes the equivalence of
the cb-norm distance and the Bures distance for completely
positive maps. We briefly discuss applications to quantum
information theory.
\end{abstract}



\end{frontmatter}




\section{Introduction and Overview}
    \label{sec:intro}

Completely positive maps (cp maps, for short) describe the dynamics of open quantum systems. Stinespring's dilation theorem \cite{Sti55,Pau02} is the basic structure theorem for such maps. It states that any cp map $T \mathpunct: \A \rightarrow \B$ between two $C^{*}$-algebras $\A$ and $\B \subset \bh$ can be written as a concatenation of two basic cp maps: a $^{*}$-homomorphism $\pi \mathpunct : \A \rightarrow \bkk$ into a larger ({\em dilated}) algebra $\bkk$ (the bounded operators on some Hilbert space $\kk$), followed by a compression $V^{*} (\cdot) V^{}$ into the range algebra $\B$:
\begin{equation}
    \label{eq:intro00}
        T(a) = V^{*} \pi(a) V^{} \qquad \forall \; \; a \in \A \, .
\end{equation}
Stinespring's theorem provides a neat characterization of the set of permissible quantum operations and is also a most useful tool in the theory of open quantum systems and quantum information \cite{NC00,Wer01}. In a way, the increased system size is the price one has to pay for a simpler description of the map $T$ in terms of just two basic operations.

A triple $(\pi, V, \kk)$ such that Eq.~(\ref{eq:intro00}) holds is called a {\em Stinespring representation} for $T$. Stinespring's representation is unique up to partial isometries on the dilation spaces: given two representations $(\pi_1, V_1, \kk_1)$ and $(\pi_2, V_2, \kk_2)$ for a completely positive map $T \mathpunct : \A
\rightarrow \bh$, there exists a partial isometry $U \mathpunct :
\kk_1 \rightarrow \kk_2$ such that
\begin{equation}
    \label{eq:intro03}
                U_{} V_{1} = V_{2} \, , \qquad U^{*}_{} V_{2} = V_{1} \, , \qquad
                {\rm and} \qquad U_{} \, \pi_{1}(a) = \pi_{2}(a) \, U_{}
\end{equation}
for all $a \in \A$. A Stinespring representation $(\pi,
V, \kk)$ of a cp map $T \mathpunct : \A \rightarrow \bh$ is called
{\em minimal} iff the set $\{ \pi(a) \, V \ket{\psi} \mid  a \in \A,
\ket{\psi} \in \hh \}$ is dense in $\kk$. If $(\pi_1, V_1, \kk_1)$ and $(\pi_2,V_2, \kk_2)$ are two minimal dilations for the cp map $T$, then
$U$ in Eq.~(\ref{eq:intro03}) is unitary. Hence, any two minimal
dilations are unitarily equivalent.

Our contribution is a continuity theorem for Stinespring's dilation: two cp maps, $T_{1}$ and $T_{2}$, are close in cb-norm iff there exist corresponding
dilations, $V_{1}$ and $V_{2}$, that are close in operator norm:
\begin{equation}
        \label{eq:intro00a}
            \frac{\cb{T_1-T_2}}{\sqrt{\cb{T_1}}+\sqrt{\cb{T_2}}}
            \leq \inf_{V_{1},V_{2}} \, \norm{V_{1} - V_{2}}
            \leq\sqrt{\cb{T_1-T_2}} \; .
\end{equation}
This result generalizes the uniqueness clause in Stinespring's theorem to cp maps that differ by a finite amount. As we have seen, uniqueness holds only up to partial isometries on the dilation spaces. So we cannot expect that any two dilations satisfy such a norm bound, only that they can be {\em chosen} in a suitable way. Hence the infimum in Eq.~(\ref{eq:intro00a}).

The {\em norm of complete boundedness} ({\em cb-norm,} for short) $\cb{\cdot}$ that appears in the continuity bound Eq.~(\ref{eq:intro00a}) is a stabilized version of the standard operator norm: For a linear map $R \mathpunct : \A \rightarrow \B$ between $C^{*}$-algebras $\A$ and $\B$, we set $\cb{R} := \sup_{n \in \N} \norm{R \otimes \id_{n}}$, where $\id_{n}$ denotes the identity map on the $(n \times n)$
matrices, and $\norm{R} := \sup_{\norm{a} \leq 1} \norm{R(a)}$.
Maps $R$ for which $\cb{R}$ is finite are
usually called {\em completely bounded}. In particular, any completely positive map $R$ is completely bounded, and we have $\cb{R} = \norm{R} =
\norm{R(\1_{\A})} = \norm{V^{*}V^{}} = \norm{V^{}}^2$, where $V$ is a Stinespring dilation for $R$. Obviously, $\norm{R} \leq
\cb{R}$ for every completely bounded map $R$. If the range algebra is Abelian, we even have equality: $\norm{R} = \cb{R}$. An Abelian domain is still enough to ensure that positive maps are completely positive, but not sufficient to guarantee that bounded maps are completely bounded \cite{Pau02}. Quantum systems typically show a separation between stabilized and unstabilized norms \cite{HLS+04,KSW06}. Hence, Eq.~(\ref{eq:intro00a}) will in general fail to hold if the cb-norm $\cb{\cdot}$ is replaced by the standard operator norm $\norm{\cdot}$.

The continuity bound Eq.~(\ref{eq:intro00a}) shows that the distance between two cp maps can equivalently be evaluated in terms of their dilations. We call this distance measure the {\em Bures distance}, since it generalizes Bures's metric \cite{Bur69} from positive functionals to general cp maps. In Sec.~\ref{sec:main} we will formally introduce the Bures distance between general cp maps and state the continuity theorem. The remainder of the article is devoted to the proof of the theorem. Sec.~\ref{sec:lower} gives the lower bound on the Bures distance in
terms of the cb-norm, which is elementary. The upper
bound is established in Sec.~\ref{sec:upper}; it relies on Bures's corresponding result for positive functionals \cite{Bur69} and on Ky Fan's minimax theorem. We first discuss cp maps with range $\bh$, and then extend the results to cp maps with injective range in Sec.~\ref{sec:general}. We
conclude with a pair of appendices: In App.~A we show that the Bures distance
is indeed a metric on the set of completely positive maps, and in
App.~B for completeness we reproduce Bures's proof of the upper bound for positive functionals.

Building on earlier work by Belavkin {\em et al.} \cite{BAR05}, the continuity theorem has appeared in \cite{KSW06} for the special case of unital cp maps (i.\,e., quantum channels) between finite-dimensional matrix algebras and has been applied to derive bounds on the tradeoff between information gain and disturbance in quantum physics, to establish a continuity bound for the no-broadcasting theorem, and
to improve security bounds for quantum key distribution with
faulty devices. A generalization to channels between direct sums
of finite-dimensional matrix algebras has been used to derive a
strengthened impossibility proof for quantum bit commitment
\cite{DKS+06}.


\section{Main Results}
    \label{sec:main}

The {\em Bures distance} evaluates the distance between two cp maps in terms of their dilations. We first discuss maps with range algebra $\B = \bh$, the bounded operators on some Hilbert space $\hh$. 
\begin{define}
    \label{def:bures01}
        {\bf (Bures Distance)}\\
         Assume a $C^{*}$-algebra $\A$, a Hilbert space $\hh$, and two cp maps  $T_{i} \mathpunct: \A \rightarrow \bh$.
        \begin{enumerate}
            \item
                 The {\em $\pi$-distance between $T_1$ and $T_2$} is defined as
                \begin{equation}
                    \label{eq:main01}
                        \beta_{\pi}(T_{1}, T_{2}) := \inf \big \{ \norm{V_{1} -
                        V_{2}} \mid V_{i} \in S(T_{i},\pi) \big \} \, ,
                \end{equation}
                where the {\em $\pi$-fiber} $S(T,\pi)$ of a cp map $T \mathpunct: \A \rightarrow \bh$ and a representation $\pi \mathpunct : \A \rightarrow \bkk$ is defined
                as the set of all operators $V \mathpunct : \hh \rightarrow \kk$ such that $(\pi, V, \kk)$ dilates $T$.
                If one or both of the fibers are empty, we set $\beta_{\pi}(T_{1},
                T_{2}) := 2$. \\
            \item
                The {\em Bures distance between $T_1$ and $T_2$} is the smallest such $\pi$-distance:
                \begin{equation}
                    \label{eq:main02}
                        \beta(T_{1}, T_{2}) := \inf_{\pi} \beta_{\pi}(T_{1},
                        T_{2}) \, ,
                \end{equation}
                with $\beta_{\pi}$ as in Eq.~(\ref{eq:main01}).
        \end{enumerate}
\end{define}

For cp maps with one-dimensional range algebra, i.e. positive functionals,
$\beta$ coincides with Bures's distance function, as introduced in
his seminal $1969$ paper \cite{Bur69}. Our definition is the natural
generalization to arbitrary cp maps; we hence choose the same
name. The statement of the continuity theorem amounts to showing that the cb-norm and the Bures distance are equivalent distance measures for cp maps. 

\begin{theo}
    \label{theo:cont01}
        {\bf (Continuity of Stinespring's Dilation)}\\
        Let $\A$ be a C*-algebra, and let $T_i \mathpunct : \A\to\bh$ be completely positive maps such that $T_i \neq 0$ for at least one $i \in \{1,2\}$. With $\beta(T_{1}, T_{2})$ defined as in Eq.~(\ref{eq:main02}), we then have the following inequality:
        \begin{equation}
            \label{eq:main03}
                \frac{\cb{T_1-T_2}}{\sqrt{\cb{T_1}}+\sqrt{\cb{T_2}}}
                \leq\beta(T_1,T_2)\leq\sqrt{\cb{T_1-T_2}} \; .
        \end{equation}
        Moreover, there exist a common representation $\pi \mathpunct : \A \to \bkk$ for $T_1$ and $T_2$ and two corresponding Stinespring dilations
        $V_i \mathpunct : \hh \to \kk$ such that
        \begin{equation}
            \label{eq:main04}
                \norm{V_{1} - V_{2}} = \beta_{\pi}(T_{1}, T_{2}) =
                \beta_{}(T_{1}, T_{2}) \, .
        \end{equation}
\end{theo}

If $(\hat{\pi}_i, \hat{V}_i, \hat{\kk}_i)$ is the minimal
Stinespring dilation for the cp map $T_i$, we can choose $\pi :=
\hat{\pi}_1 \oplus \hat{\pi}_2$ as the common
representation in Th.~\ref{theo:cont01}.
Even more is known for positive functionals: in that case the Bures
distance can be evaluated in {\em any} common representation
\cite{Ara72,Alb83, AP00}. We do not yet know whether this result
extends to general cp maps.

What about general range algebras $\B \neq \bh$? Since any $C^{*}$-algebra $\B$ can be faithfully embedded into a norm-closed self-adjoint algebra $\bh$ with a suitably chosen Hilbert space $\hh$, it may appear natural to define the Bures distance for cp maps $T_{i} \mathpunct : \A \rightarrow \B$ in terms of the concatenated maps $\sigma_{} \circ T_{i}$, with a faithful representation $\sigma \mathpunct : \B \rightarrow \bh$. However, $\beta( \sigma_{} \circ T_{1}, \sigma_{} \circ T_{2})$ might possibly depend on the embedding representation $\sigma$. We instead choose an intrinsic definition of the Bures distance --- and show that it reduces to Def.~\ref{def:bures01} if $\B = \bh$.

\begin{define}
    \label{def:bures02}
        {\bf (Bures Distance for General Range Algebras)}\\
        Given two $C^{*}$-algebras $\A$ and $\B$ and two cp maps  $T_{i} \mathpunct: \A \rightarrow \B$, the {\em Bures distance} is defined as
        \begin{equation}
            \label{eq:main02a}
                \beta(T_{1}, T_{2}) := \inf_{\hat{T}} \, \norm{\hat{T}_{11}(\1_{\A}) + \hat{T}_{22}(\1_{\A})- \hat{T}_{12}(\1_{\A}) - \hat{T}_{21}(\1_{\A})}^{\frac{1}{2}} \, .
        \end{equation}
        The infimum in Eq.~(\ref{eq:main02a}) is taken over all completely positive extensions $\hat{T} \mathpunct : \A \rightarrow \B \otimes \bc{2} \simeq \M_{2} (\B)$ of the form
        \begin{equation}
            \label{eq:main02b}
                \hat{T} \simeq \left ( \begin{array}{cc} \hat{T}_{11} \; & \hat{T}_{12} \; \\ \hat{T}_{21} \; & \hat{T}_{22} \; \end{array} \right )
        \end{equation}
        with completely bounded maps $\hat{T}_{ij} \mathpunct : \A \rightarrow \B$ satisfying $\hat{T}_{ii} = T_{i}$.
\end{define}

Introducing the cp map $\eta \mathpunct : \bc{2} \rightarrow \C$ by setting $\eta(x) := {\rm tr} \, {\scriptsize \left ( \begin{array}{cc} 1 & -1 \\ -1 & 1 \end{array} \right )} x$, Eq.~(\ref{eq:main02a}) can be rewritten more compactly,
\begin{equation}
    \label{eq:main02c}
        \beta(T_{1}, T_{2}) = \inf_{\hat{T}} \, \norm{ (\id_{\B} \otimes \eta_{}) \circ \hat{T}_{}(\1_{A})}^{\frac{1}{2}} \, .
\end{equation}
While this definition of the Bures distance admittedly looks quite different from Def.~\ref{def:bures01}, we will show in Sec.~\ref{sec:consistent} that the definitions coincide if $\B = \bh$, and hence it is justified to use the same symbol for both.

With this definition of the Bures distance, Th.~\ref{theo:cont01} can now be generalized to cp maps with injective range algebras. Recall that a $C^{*}$-algebra $\B$ is called {\em injective} if for every $C^{*}$-algebra $\A$ and operator system $\calS$ contained in $\A$, every completely positive map $R \mathpunct : \calS \rightarrow \B$ can be extended to a completely positive map on all of $\A$ (cf. \cite{Pau02}, Ch.~7). In fact, in order to show that $\B \subset \bh$ is injective it is enough to find a completely positive map $P \mathpunct : \bh \rightarrow \B$ such that $P(b) = b$ for all $b \in \B$. $P$ is usually called a {\em completely positive conditional expectation.} Connes has shown that a von Neumann algebra $\B$ is injective iff it is {\em hyperfinite,} which means that $\B$ contains an ascending sequence of finite dimensional subalgebras with dense union. We refer to Ch.~XVI in Takesaki's textbook \cite{Tak03} for this and further equivalent conditions for injectivity of von Neumann algebras. A characterization of injective $C^{*}$-algebras has been given by Robertson {\em et al.} \cite{RW89,RY90}. For cp maps with non-injective range, we only have a lower bound on $\beta(T_{1},T_{2})$, though we could always apply Th.~\ref{theo:cont01} to the concatenated maps $\sigma \circ T_{1}$ with some faithful embedding $\sigma$.

\begin{theo}
    \label{theo:cont02}
        {\bf (Continuity for General Range Algebras)}\\
        Let $\A$ and $\B$ be $C^{*}$-algebras, and let $T_{i} \mathpunct : \A \rightarrow \B$ be completely positive. With $\beta(T_{1}, T_{2})$ defined as in Def.~\ref{def:bures02} above, we have
        \begin{equation}
            \label{eq:main05}
                \frac{\cb{T_1-T_2}}{\sqrt{\cb{T_1}}+\sqrt{\cb{T_2}}}
                \leq\beta(T_1,T_2) \, .
        \end{equation}
        If in addition $\B$ is injective, we also have
        \begin{equation}
            \label{eq:main06}
                \beta(T_1,T_2)\leq\sqrt{\cb{T_1-T_2}} \; ,
        \end{equation}
        and $\beta(T_{1}, T_{2}) = \beta (\sigma_{} \circ T_{1}, \sigma_{} \circ T_{2})$ for any faithful representation $\sigma \mathpunct : \B \rightarrow \bh$.
\end{theo}

The remainder of the article is devoted to the proof of Theorems~\ref{theo:cont01} and \ref{theo:cont02}. We start in Sec.~\ref{sec:lower} with a lower bound on the Bures distance in
terms of the cb-norm.


\section{Lower Bound}
    \label{sec:lower}

 A lower bound on the Bures distance $\beta(T_{1}, T_{2})$ in terms of the cb-norm distance $\cb{T_{1} - T_{2}}$ easily follows from the standard properties of the operator norm.

 \begin{propo}
    \label{propo:lower}
        {\bf (Lower Bound)}\\
        Let $\A$ be a $C^{*}$-algebra, and $T_{1}, T_{2} \mathpunct :
        \A \rightarrow \bh$ be completely positive maps. We then have
        \begin{equation}
            \label{eq:lower01}
                \cb{T_{1}-T_{2}} \leq \big ( \sqrt{\cb{T_{1}}} + \sqrt{\cb{T_{2}}} \, \big ) \, \beta(T_{1}, T_{2}) \, .
        \end{equation}
 \end{propo}
{\bf Proof:} Let $\pi$ be a common representation for the cp maps $T_{i}$ with corresponding dilations $(\pi_{}, V_{i}, \kk_{})$. Given $n \in \N$ and $x \in \A \otimes \bc{n}$, we can then apply the triangle inequality to conclude that
\begin{equation}
    \label{eq:lower02}
        \begin{split}
            \norm{T_{1} & \otimes \id_{n} (x) - T_{2} \otimes \id_{n} (x)}\\
             & = \norm{(V_{1}^{*} \otimes \1_{n}) \, (\pi_{} \otimes \id_{n}) (x) \, (V_{1}^{} \otimes \1_{n}^{}) - (V_{2}^{*} \otimes \1_{n}) \, (\pi_{} \otimes \id_{n}) (x) \, (V_{2}^{} \otimes \1_{n}^{})}\\
             & \leq \norm{ \big ( ( V_{1}^{*} - V_{2}^{*}) \otimes \1_{n}^{} \big ) \, \big ( \pi_{} \otimes \id_{n} \big)(x) \, (V_{1}^{} \otimes \1_{n}^{})} \\
             & \qquad \qquad + \norm{ (V_{2}^{*} \otimes \1_{n}) \, (\pi_{} \otimes \id_{n}) (x) \, \big ( ( V_{1} - V_{2}) \otimes \1_{n}\big)}\\
             & \leq \norm{V_{1} - V_{2}} \, \norm{V_{1}} \, \norm{x} + \norm{V_{1} - V_{2}} \, \norm{V_{2}} \, \norm{x}\\
             & = \big ( \sqrt{\cb{T_{1}}} + \sqrt{\cb{T_{2}}} \, \big ) \, \norm{V_{1} - V_{2}} \, \norm{x} \, ,
            \
        \end{split}
\end{equation}
where we have used that the operator norm is preserved under both the adjoint operation and tensoring with the identity $\1_n$, as well as $\cb{\pi} = 1$. The statement then immediately follows from the definition of the cb-norm and the Bures distance. $\blacksquare$


\section{Upper Bound}
    \label{sec:upper}

In this Section we will complement Prop.~\ref{propo:lower} with an upper bound on the Bures
distance $\beta(T_{1}, T_{2})$ in terms of the cb-norm $\cb{T_{1}
- T_{2}}$. We start by investigating several alternative ways to evaluate
the Bures distance --- a useful tool for our proof but also a result of independent interest.

Given two cp maps $T_{i} \mathpunct : \A
\rightarrow \bh$ and a representation $\pi \mathpunct : \A
\rightarrow \bkk$, we set
\begin{equation}
    \label{eq:upper01}
        \n_{\pi}(T_{1},T_{2}) := \{ V_{1}^{*} V_{2}^{} \mid V_{i}
        \in S(T_{i},\pi) \} \subset \bh \, .
\end{equation}
The $\pi$-distance $\beta_{\pi}(T_{1},T_{2})$ can now be calculated
in terms of $\n_{\pi}(T_{1}, T_{2})$ as follows:

\begin{lemma}
    \label{lemma:N}
        For cp maps $T_{i} \mathpunct : \A
        \rightarrow \bh$ and a representation $\pi \mathpunct : \A
        \rightarrow \bkk$, we have
        \begin{equation}
            \label{eq:upper02}
                \beta_{\pi}^{2}(T_{1}, T_{2}) = \inf_{N \in
                \n_{\pi}(T_{1},T_{2})} \sup_{\varrho \in \sbhstar}
                \big \{ \trace{ \varrho \, T_{1}(\1_{\A})} \, + \, \trace{ \varrho \,
                T_{2}(\1_{\A})} \, - \, 2 \, \re{{\rm tr} \,
                \varrho \, N } \big \} \, ,
        \end{equation}
        where $\sbhstar$ denotes the positive trace class operators of unit trace on the Hilbert space $\hh$.
\end{lemma}
{\bf Proof:} The map $x \mapsto \tr{(\cdot) x}$ defines an
isometric isomorphism from $\bh$ to the normalized trace class
operators $\mathcal{B}_{*,1}\mathcal{(\hh)}$ (cf. Sec.~VI.6 in
\cite{RS80}). Since in addition $(V_{1} - V_{2})^{*} (V_{1} -
V_{2})$ is positive, we can write
\begin{equation}
    \label{eq:upper03}
        \begin{split}
            \norm{V_{1} - V_{2}}^{2} & = \norm{(V_{1}^{} - V_{2}^{})^{*}
            (V_{1}^{} - V_{2}^{})}\\
            & = \sup_{\varrho \in \sbhstar} \trace{ \varrho \,
            ( V_{1} - V_{2})^{*} (V_{1} - V_{2} )} \\
            & = \sup_{\varrho \in \sbhstar}
            \big \{ \trace{ \varrho \, T_{1}(\1_{\A})} \, + \, \trace{ \varrho \,
            T_{2}(\1_{\A})} \, - \, 2 \, \re{{\rm tr} \,
            \varrho \, V_{1}^{*} V_{2}^{} } \big \}
        \end{split}
\end{equation}
for $V_{i} \in S(T_{i},\pi)$ and any given representation $\pi$.
The result then immediately follows from the definition of
$\beta_{\pi}(T_{1}, T_{2})$ in Eq.~(\ref{eq:main01}) and
$\n_{\pi}(T_{1},T_{2})$ in Eq.~(\ref{eq:upper01}). $\blacksquare$

The following lemma allows to replace the infimum over representations $\pi$ and corresponding $N \in \n_{\pi}$ in Lemma~\ref{lemma:N} with an infimum over intertwiners $W \mathpunct : \kk_{2} \rightarrow \kk_{1}$ between any two fixed Stinespring representations. As advertised in Sec.~\ref{sec:main}, we will also show how to find a common
representation $\pi$ such that $\beta_{}(T_{1},T_{2}) =
\beta_{\pi}(T_{1},T_{2})$.

\begin{lemma}
    \label{lemma:evaluation}
    {\bf (Evaluation of the Bures Distance)}\\
    Let $\A$ be a $C^{*}$-algebra, $\hh$ a Hilbert space, and
    $T_{1}, T_{2} \mathpunct : \A \rightarrow \bh$ be two
    completely positive maps.
    \begin{enumerate}
        \item Assuming Stinespring dilations $(\pi_{i}, V_{i},
        \kk_{i})$ for $T_{i}$, we define
        \begin{equation}
            \label{eq:upper04}
                \M(T_{1},T_{2}):= \big \{ V_{1}^{*} W V_{2}^{} \mid  W \pi_{2}(a) =
                \pi_{1}(a) W  \; \forall \; a \in \A \, , \norm{W} \leq 1 \big
                \} \, .
        \end{equation}
        The set $\M(T_{1},T_{2}) \subset \bh$ depends only on the
        cp maps $T_{i}$, not on the dilations $(\pi_{i}, V_{i},
        \kk_{i})$.\\
        \item The set $\M(T_{1}, T_{2})$ can be represented
        alternatively as
        \begin{equation}
            \label{eq:upper05}
                \M(T_{1}, T_{2}) = \bigcup_{\pi} \n_{\pi}(T_{1},T_{2}) =:
                \n(T_{1}, T_{2}) \, ,
        \end{equation}
        where the union is over all representations $\pi$
        admitting a common Stinespring representation for $T_{1}$
        and $T_{2}$, and $\n_{\pi}(T_{1}, T_{2})$
        is defined in Eq.~(\ref{eq:upper01}).\\
        \item There exists a representation $\pi$ such that
        $\beta_{} (T_{1}, T_{2}) = \beta_{\pi} (T_{1}, T_{2})$.
        We can choose $\pi := \hat{\pi}_{1} \oplus \hat{\pi}_{2}$,
        where $\hat{\pi}_{i}$ is a minimal representation for
        $T_{i}$.
    \end{enumerate}
\end{lemma}
{\bf Proof:} $(1)$ For the first part, our strategy is to show that $\M(T_{1}, T_{2})$,
defined via some dilations $(\pi_{i}, V_{i},\kk_{i})$, coincides
with $\hat{\M}(T_{1}, T_{2})$ defined via the minimal dilations
$(\hat{\pi}_{i}, \hat{V}_{i},\hat{\kk}_{i})$. Given two dilations
$(\pi_{i}, V_{i},\kk_{i})$ for $T_{1}$ and $T_{2}$, respectively,
we know from the uniqueness clause in Stinespring's theorem that there exist isometries
$U_{i} \mathpunct : \hat{\kk}_{i} \rightarrow \kk_{i}$ such that
$U_{i} \hat{V}_{i} = V_{i}$ and $U_{i}^{*} V_{i}^{} =
\hat{V}_{i}$. Since $U_{i}^{} U_{i}^{*}$ is a projector onto the
closed linear span of $\{\pi_{i}(a) V_{i} \ket{\psi} \}$, we have
$U_{i}^{} U_{i}^{*} V_{i} = V_{i}$, and hence
\begin{equation}
    \label{eq:upper06}
        V_{1}^{*} \, W \, V_{2}^{} = V_{1}^{*} \, U_{1}^{} \,
        U_{1}^{*} \, W \, U_{2}^{} \, U_{2}^{*} \,
        V_{2}^{} \, = \, \hat{V}_{1}^{*} \, U_{1}^{*} \, W \, U_{2}^{} \,
        \hat{V}_{2}^{} = \hat{V}_{1}^{*} \, \hat{W} \,
        \hat{V}_{2}^{}
\end{equation}
for all $W \mathpunct : \kk_{2} \rightarrow \kk_{1}$, where we
have set $\hat{W} := U_{1}^{*} W U_{2}^{} \mathpunct :
\hat{\kk}_{2} \rightarrow \hat{\kk}_{1}$. The intertwining
relations $U_{i} \, \hat{\pi}_{i}(a) = \pi_{i}(a) \, U_{i}$ and
$W\, \pi_{2}(a) = \pi_{1}(a) \, W$ imply that
\begin{equation}
    \label{eq:upper07}
    \begin{split}
        \hat{W} \, \hat{\pi}_{2}(a) & = U_{1}^{*} \, W \, U_{2}^{}
        \, \hat{\pi}_{2}(a)\\
        & = U_{1}^{*} \, W \, \pi_{2}(a) \, U_{2}\\
        & = U_{1}^{*} \, \pi_{1}(a) \, W \, U_{2}^{} \\
        & = \hat{\pi}_{1}(a) \, U_{1}^{*} \, W \, U_{2}^{}\\
        & = \hat{\pi}_{1}(a) \, \hat{W}
    \end{split}
\end{equation}
for all $a \in \A$. Moreover, $\norm{\hat{W}} = \norm{U_{1}^{*} W
U_{2}^{}}\leq \norm{W} \leq 1$, since the $U_{i}$ are isometric.
Hence, $\M(T_{1}, T_{2}) \subset \hat{\M}(T_{1},T_{2})$. The
converse is completely analogous, starting with $\hat{W}$ and
setting $W := U_{1}^{} \hat{W} U_{2}^{*}$. $\blacktriangle$

$(2)$ In order to show that $\M(T_{1}, T_{2}) \subset \n (T_{1}, T_{2})$, it is sufficient to find a common representation $\pi$ such that $\M_{}(T_{1}, T_{2}) \subset \n_{\pi}(T_{1}, T_{2})$. Since
$\M(T_{1}, T_{2})$ is independent of the dilations according to part $(1)$, we can assume it to be defined via the minimal dilations $(\hat{\pi}_{i},
\hat{V}_{i},\hat{\kk}_{i})$. Given $\hat{W} \mathpunct :
\hat{\kk}_{2} \rightarrow \hat{\kk}_{1}$ such that $\norm{\hat{W}}
\leq 1$, we define the bounded operators $V_{i} \mathpunct : \hh
\rightarrow \hat{\kk}_{1} \oplus \hat{\kk}_{2}$ by setting
\begin{align}
    \label{eq:upper08}
            V_{1} \ket{\psi} & := \hat{V}_{1} \ket{\psi} \; \oplus \; 0 \, ,\\
    \label{eq:upper08a}
            V_{2} \ket{\psi} & := \hat{W} \, \hat{V}_{2} \ket{\psi} \;
            \oplus \; \sqrt{\1_{\hat{\kk}_{2}} -
            \hat{W}^{*} \hat{W}^{}} \, \hat{V}_{2} \ket{\psi} \, .
\end{align}
Making use of the intertwining relation $\hat{W} \hat{\pi}_{2}(a) =
\hat{\pi}_{1}(a)\hat{W}$, it is then
straightforward to verify that $\hat{\pi}_{1} \oplus
\hat{\pi}_{2}$ is indeed a common representation for the cp maps $T_{1}$
and $T_{2}$, with Stinespring dilations
$(\hat{\pi}_{1} \oplus \hat{\pi}_{2}, V_{i}, \hat{\kk}_{1} \oplus
\hat{\kk}_{2})$. Moreover, $\hat{V}_{1}^{*} \hat{W} \hat{V}_{2} =
V_{1}^{*}V_{2}^{} \in \n_{\hat{\pi}_{1} \oplus
\hat{\pi}_{2}}(T_{1},T_{2}) \subset \n_{} (T_{1}, T_{2})$, as
suggested. In particular, the direct sum construction shows that
we can always find a common representation for the cp maps
$T_{i}$, and hence $\n(T_{1},T_{2})$ is always non-empty. For the
converse implication, $\n(T_{1},T_{2}) \subset \M(T_{1},T_{2})$, let
$\pi$ be any such common representation and $V_{1}^{*} V_{2}^{}
\in \n_{\pi}(T_{1},T_{2})$. Defining $\M (T_{1}, T_{2})$ via the
dilations $(\pi_{}, V_{i}, \kk_{})$ and choosing $W_{}=\1_{\kk}$, we have
$\n_{\pi}(T_{1},T_{2}) \subset \M_{}(T_{1},T_{2})$.
$\blacktriangle$

$(3)$  From the proof of part $(2)$ we have $\n_{\pi} (T_{1},T_{2}) \subset \M_{} (T_{1}, T_{2}) \subset \n_{\hat{\pi}_{1} \oplus \hat{\pi}_{2}} (T_{1}, T_{2})$ for any common representation $\pi$. We can then immediately conclude
from Lemma~\ref{lemma:N} that
\begin{equation}
    \label{eq:eq:upper11}
        \beta_{\hat{\pi}_{1} \oplus \hat{\pi}_{2}} (T_{1}, T_{2})
        \leq \beta_{\pi} (T_{1},T_{2}) \, ,
\end{equation}
implying $\beta_{}(T_{1},T_{2}) = \beta_{\hat{\pi}_{1} \oplus
\hat{\pi}_{2}} (T_{1}, T_{2})$. Consequently, the Bures distance
can always be evaluated in the direct sum representation of the
minimal representations. $\blacksquare$

Lemma~\ref{lemma:N} and Lemma~\ref{lemma:evaluation} can now be applied to derive the desired upper bound on the Bures distance in
terms of the cb-norm. For the special case
of positive functionals, this result was obtained by Bures \cite{Bur69}
(cf. Prop.~\ref{propo:bures} in App.~B), and will now be lifted to
cp maps with the help of Ky Fan's minimax theorem \cite{Fan53}.

\begin{propo}
    \label{propo:upper}
        {\bf (Upper Bound)}\\
        Let $\A$ be a $C^{*}$-algebra, and $T_{1}, T_{2} \mathpunct :
        \A \rightarrow \bh$ be completely positive maps. We can
        then find a common representation $\pi \mathpunct : \A \rightarrow \bkk$ and
        corresponding dilations $(\pi_{}, V_{i}, \kk_{})$ for $T_{i}$ such
        that
        \begin{equation}
            \label{eq:upper12}
                \norm{V_{1}-V_{2}} = \beta_{\pi}(T_{1}, T_{2}) = \beta_{}(T_{1},
                T_{2}) \leq \sqrt{\cb{T_{1} -T_{2}}} \; .
        \end{equation}
\end{propo}
{\bf Proof:} Spelling out $\beta_{\pi}(T_{1},T_{2})$ as in
Lemma~\ref{lemma:N} and then making use of the relation $\n(T_{1},
T_{2}) = \M(T_{1}, T_{2})$ from Lemma~\ref{lemma:evaluation}, we
have
\begin{equation}
    \label{eq:upper13}
        \begin{split}
            \beta^{2}_{}(T_{1},T_{2}) & = \inf_{\pi}
            \beta_{\pi}^{2}(T_{1},T_{2}) \\
            & = \inf_{N \in \n(T_{1}, T_{2})} \sup_{\varrho \in \sbhstar}
            \big \{ \trace{ \varrho \, T_{1}(\1_{\A})} \, + \, \trace{ \varrho \,
            T_{2}(\1_{\A})} \, - \, 2 \, \re{{\rm tr} \,
            \varrho \, N } \big \} \\
            & = \inf_{M \in \M(T_{1}, T_{2})} \sup_{\varrho \in \sbhstar}
            \big \{ \trace{ \varrho \, T_{1}(\1_{\A})} \, + \, \trace{ \varrho \,
            T_{2}(\1_{\A})} \, - \, 2 \, \re{{\rm tr} \,
            \varrho \, M } \big \} \\
            & = \inf_{\norm{W}\leq 1} \sup_{\varrho \in \sbhstar}
            \big \{ \trace{ \varrho \, T_{1}(\1_{\A})} \, + \, \trace{ \varrho \,
            T_{2}(\1_{\A})} \, - \, 2 \, \re{{\rm tr} \,
            \varrho \, V_{1}^{*} W V_{2}^{} } \big \}
        \end{split}
\end{equation}
with $W \in \mathcal{B}(\kk_{2}, \kk_{1})$, where $(\pi_{1}, V_{1}, \kk_{1})$ and $(\pi_{2}, V_{2}, \kk_{2})$
are now any two {\em fixed} dilations for the cp maps $T_{1}$ and
$T_{2}$, respectively. The target functional in
Eq.~(\ref{eq:upper13}) is affine in both inputs. Since the state
$\varrho \in \sbhstar$ is trace-class, so is $V_{2}^{} \varrho
V_{1}^{*}$, and hence the functional is weakly continuous in $W$.
Moreover, we know from the
Banach-Alaoglu theorem (cf. Sec.~IV.5 in \cite{RS80}) that the
unit ball $\norm{W} \leq 1$ is weakly compact, and hence the
infimum is attained. In addition, both optimizations in
Eq.~(\ref{eq:upper13}) are performed over convex sets. Under these
conditions, Ky Fan's minimax theorem \cite{Fan53,Sim98} guarantees
that the order of the optimizations in Eq.~(\ref{eq:upper13}) can
be interchanged to yield
\begin{equation}
    \label{eq:upper14}
        \begin{split}
            \beta^{2}_{}(T_{1},T_{2}) & = \min_{\norm{W}\leq 1}
            \sup_{\varrho \in \sbhstar} \big \{ \trace{ \varrho
            \, T_{1}(\1_{\A})} \, + \, \trace{ \varrho \,
            T_{2}(\1_{\A})} \, - \, 2 \, \re{{\rm tr} \,
            \varrho \, V_{1}^{*} W V_{2}^{} } \big \} \\
            & = \sup_{\varrho \in \sbhstar}
            \min_{\norm{W}\leq 1} \big \{ \trace{ \varrho
            \, T_{1}(\1_{\A})} \, + \, \trace{ \varrho \,
            T_{2}(\1_{\A})} \, - \, 2 \, \re{{\rm tr} \,
            \varrho \, V_{1}^{*} W V_{2}^{} } \big \} \\
            & = \sup_{\ket{\psi} \in \hh \otimes \hh}
            \min_{\norm{W}\leq 1} \big \{ \bra{\psi}
            T_{1}(\1_{\A}) \otimes \1_{\hh} \ket{\psi} \,
            + \, \bra{\psi} T_{2}(\1_{\A})
            \otimes \1_{\hh} \ket{\psi}\\
            & \qquad \qquad - \, 2 \, \re{
            \bra{\psi} ( V_{1}^{*} \otimes \1_{\hh}) ( W \otimes
            \1_{\hh} ) (V_{2}^{} \otimes \1_{\hh}) \ket{\psi}}
            \big \} \, .
        \end{split}
\end{equation}
In the last step of Eq.~(\ref{eq:upper14}), we have replaced the supremum over the normal
states $\varrho \in \sbhstar$ by a supremum over their
respective purifications. Note that $(\pi_{i} \otimes \id_{\bh}, (V_{i}
\otimes \1_{\hh}) \ket{\psi} , \kk_{i} \otimes \hh)$ is a Stinespring dilation
for the positive functional $\psi \circ (T_{i} \otimes
\id_{\bh})$, and that all operators $\tilde{W} \mathpunct :
\kk_{2} \otimes \hh_{} \rightarrow \kk_{1} \otimes \hh_{}$ that
intertwine the representations $\pi_{1} \otimes \id_{\bh}$ and
$\pi_{2} \otimes \id_{\bh}$ are of the form $\tilde{W} = W \otimes
\1_{\hh}$, with an intertwiner $W \mathpunct : \kk_{2} \rightarrow
\kk_{1}$. Lemma~\ref{lemma:evaluation} therefore implies that the inner
variation in Eq.~(\ref{eq:upper14}) is just the Bures distance
square $\beta^{2} \big ( \psi \circ (T_{1} \otimes \id_{\bh}),
\psi \circ (T_{2} \otimes \id_{\bh}) \big )$. We can then apply
Bures's bound for positive functionals from Prop.~\ref{propo:bures}
to conclude that
\begin{equation}
    \label{eq:upper15}
        \begin{split}
            \beta^{2}_{}(T_{1},T_{2}) & = \sup_{\ket{\psi} \in \hh \otimes
            \hh} \beta^{2} \big ( \psi \circ (T_{1} \otimes \id_{\bh}),
            \psi \circ (T_{2} \otimes \id_{\bh}) \big )\\
            & \leq \sup_{\ket{\psi} \in \hh \otimes \hh}
            \norm{\psi \circ (T_{1} \otimes
            \id_{\bh}) - \psi \circ (T_{2} \otimes \id_{\bh})}\\
            & \leq \cb{T_{1} - T_{2}} \, ,
        \end{split}
\end{equation}
which is the desired result. For the cb-norm bound in the last step we have used that the finite rank operators are dense in $\bhstar$. We have seen above that there exists an intertwiner $W \mathpunct : \kk_{2} \rightarrow \kk_{1}$
which attains the infima in Eq.~(\ref{eq:upper13}) and
Eq.~(\ref{eq:upper14}). Lemma~\ref{lemma:evaluation} then by
construction yields a common representation $\pi$ and
corresponding dilations $(\pi, V_{i}, \kk)$ such that
\begin{equation}
    \label{eq:upper16}
        \norm{V_{1}-V_{2}} = \beta(T_{1}, T_{2}) \leq \sqrt{ \cb{T_{1} - T_{2}}}
        \; ,
\end{equation}
just as claimed. $\blacksquare$

Th.~\ref{theo:cont01} now immediately follows by combining the bounds from Prop.~\ref{propo:lower} and Prop.~\ref{propo:upper}.


\section{Bures Distance for General Range Algebras}
    \label{sec:general}

So far our discussion has focused on channels with range algebra $\bh$. In this Section we will investigate completely positive maps $T_{i} \mathpunct : \A \rightarrow \B$ with general range algebra $\B$. Our results are twofold: in Sec.~\ref{sec:consistent} we will justify the intrinsic definition of the Bures distance $\beta(T_{1}, T_{2})$ by showing that it indeed coincides with Def.~\ref{def:bures01} if $\B = \bh$. For general range algebras $\B$, we will then show in Sec.~\ref{sec:injective} that $\beta(T_{1}, T_{2}) \geq \beta (\sigma_{} \circ T_{1}, \sigma_{} \circ T_{2})$ for any representation $\sigma$. If $\B$ is injective and $\sigma$ is faithful, we even have equality, and hence the Bures distance does not depend on the details of the embedding and can then be shown to be completely equivalent to the cb-norm distance. For the proof we need a monotonicity result for the Bures distance, which we will present in Sec.~\ref{sec:mono}.


\subsection{Consistency}
    \label{sec:consistent}

For the moment, we will denote the Bures distance for cp maps $T_{i} \mathpunct : \A \rightarrow \B$ with general range algebra $\B$, as introduced in Def.~\ref{def:bures02}, by $\beta'(T_{1}, T_{2})$. We will show in this Section that indeed $\beta'(T_{1}, T_{2}) = \beta(T_{1}, T_{2})$ if $\B = \bh$. Thus, Def.~\ref{def:bures02} is a consistent generalization of Def.~\ref{def:bures01} to general range algebras, and we may henceforth drop the prime.

\begin{propo}
    \label{propo:consistency}
    Let $\A$ be a $C^{*}$-algebra, and let $T_{i} \mathpunct : \A \rightarrow \bh$ be completely positive. With $\beta(T_{1}, T_{2})$ defined as in Def.~\ref{def:bures01} and $\beta'(T_{1}, T_{2})$ defined as in Def.~\ref{def:bures02}, we then have
    \begin{equation}
        \label{eq:consistent01}
            \beta(T_{1}, T_{2}) = \beta'(T_{1}, T_{2}) \, .
    \end{equation}
\end{propo}
{\bf Proof:} We first show that $\beta(T_{1}, T_{2}) \leq \beta' (T_{1}, T_{2})$. As in Def.~\ref{def:bures02}, let $\hat{T} \mathpunct : \A \rightarrow \bh \otimes \bc{2}$ be a completely positive extension of the cp maps $T_{i}$ with Stinespring dilation $(\pi, V, \kk)$. Starting from $V \mathpunct : \hh \otimes \C^2 \rightarrow \kk$, for $i \in \{ 1,2 \}$ we define $V_{i} \mathpunct : \hh \rightarrow \kk$ by setting $V_{i} \ket{\psi} := V_{} \ket{\psi} \otimes \ket{i}$. Hence, $\hat{T}_{ij}(a) = V_{i}^{*} \, \pi(a) \, V_{j}^{}$ for all $a \in \A$. In particular, $(\pi_{}, V_{i}, \kk_{})$ dilates $T_{i}$. We may then conclude from Def.~\ref{def:bures01} that
\begin{equation}
    \label{eq:consistent02}
        \begin{split}
            \beta(T_{1}, T_{2}) & \leq \norm{V_{1} - V_{2}}\\
             & = \norm{V_{1}^{*} V_{1} + V_{2}^{*} V_{2} - V_{1}^{*} V_{2} - V_{2}^{*} V_{1} }^{\frac{1}{2}}\\
             & = \norm{( \id_{\bh} \otimes \eta_{}) \circ \hat{T}(\1_{\A})}^{\frac{1}{2}}
        \end{split}
\end{equation}
holds independently of $\hat{T}$, and hence  $\beta(T_{1}, T_{2}) \leq \beta' (T_{1}, T_{2})$ follows immediately from Def.~\ref{def:bures02}. $\blacktriangle$

Conversely, we assume a common representation $\pi \mathpunct : \A \rightarrow \bkk$ for the cp maps $T_{i}$ with corresponding dilations $(\pi_{}, V_{i}, \kk)$. We now set $V_{} \, \ket{\psi} \otimes \ket{i} := V_{i} \ket{\psi}$. The linear map $V \mathpunct : \hh \otimes \C^{2} \rightarrow \kk$ defines a completely positive extension $\hat{T}(a) = V^{*} \, \pi (a) \, V^{}$ in the sense of Def.~\ref{def:bures01} with $\hat{T}_{ij}(a) = V_{i}^{*} \, \pi(a) \, V_{j}^{}$ for all $a \in \A$. Hence,
\begin{equation}
    \label{eq:consistent03}
         \begin{split}
            \beta'(T_{1}, T_{2}) & \leq \norm{( \id_{\bh} \otimes \eta_{}) \circ \hat{T}(\1_{\A})}^{\frac{1}{2}}\\
            & = \norm{V_{1}^{*} V_{1} + V_{2}^{*} V_{2} - V_{1}^{*} V_{2} - V_{2}^{*} V_{1} }^{\frac{1}{2}}\\
            & = \norm{V_{1} - V_{2}} \, ,
        \end{split}
\end{equation}
implying that $\beta'(T_{1}, T_{2}) \leq \beta(T_{1}, T_{2})$. $\blacksquare$


\subsection{Monotonicity of the Bures Distance under Cp Maps}
    \label{sec:mono}

We will now show that the Bures distance $\beta(T_{1}, T_{2})$ decreases under quantum operations. Only Eq.~(\ref{eq:mono01}) is needed in the proof of Th.~\ref{theo:cont02} below, but we include Eq.~(\ref{eq:mono02}) for completeness.

\begin{propo}
    \label{propo:mono}
    {\bf (Monotonicity)}\\
    Given three $C^{*}$-algebras $\A$, $\B$, and $\D$ and cp maps  $T_{1}, T_{2} \mathpunct : \A \rightarrow \B$ and $S \mathpunct : \B \rightarrow \D$, we have
    \begin{equation}
            \label{eq:mono01}
                \beta( S_{} \circ T_{1}, S_{} \circ T_{2}) \leq \sqrt{\norm{S_{}}} \; \beta(T_{1}, T_{2}) \, .
    \end{equation}
    For cp maps $T_{i}$ as above and $S \mathpunct : \D \rightarrow \A$ we have
    \begin{equation}
        \label{eq:mono02}
                \beta( T_{1} \circ S_{}, T_{2} \circ S_{}) \leq  \sqrt{\norm{S_{}}} \; \beta(T_{1}, T_{2}) \, .
    \end{equation}
\end{propo}
{\bf Proof:} This is straightforward. Starting with an extension $\hat{T} \mathpunct : \A \rightarrow \B \otimes \bc{2}$ for the cp maps $T_{i}$, $(S_{} \otimes \id_{2}) \circ \hat{T}$ defines a completely positive extension for the maps $S_{} \circ T_{i}$, and we have the estimate
\begin{equation}
    \label{eq:mono03}
        \beta(S_{} \circ T_{1}, S_{} \circ T_{2}) \leq \norm{ (\id_{\D} \otimes \eta) \circ (S_{} \otimes \id_{2}) \circ \hat{T} (\1_{\A})}^{\frac{1}{2}} \leq \sqrt{\norm{S}} \, \norm{(\id_{\B} \otimes \eta) \circ \hat{T} (\1_{\A})}^{\frac{1}{2}} \, .
\end{equation}
Since Eq.~(\ref{eq:mono03}) holds for all extensions $\hat{T}$, Eq.~(\ref{eq:mono01}) is proven. The proof of Eq.~(\ref{eq:mono02}) is completely analogous. $\blacksquare$


\subsection{Equivalence of Bures Distance and Cb-Norm for Injective Range Algebras}
    \label{sec:injective}

The following proposition shows that for cp maps $T_{i} \mathpunct : \A \rightarrow \B$ with injective range algebra $\B$, the Bures distance $\beta(T_{1}, T_{2})$ may be evaluated in any faithful representation $\sigma \mathpunct : \B \rightarrow \bh$.

\begin{propo}
    \label{propo:injective}
    Let $\A$ and $\B$ be $C^{*}$-algebras, and $T_{1}, T_{2} \mathpunct : \A \rightarrow \B$ be completely positive maps. We then have
    \begin{equation}
        \label{eq:injective01}
            \beta(T_{1}, T_{2}) \geq \beta (\sigma \circ T_{1}, \sigma \circ T_{2})
    \end{equation}
    for any representation $\sigma \mathpunct : \B \rightarrow \bh$. Moreover, if $\B$ is injective and the representation $\sigma$ is faithful equality holds in Eq.~(\ref{eq:injective01}).
\end{propo}
{\bf Proof:} Since $\norm{\sigma} = \cb{\sigma} = 1$ for any representation $\sigma$, Eq.~(\ref{eq:injective01}) is immediate from Prop.~\ref{propo:mono}. For the converse inequality, assume that $\B$ is injective and $\sigma \mathpunct : \B \rightarrow \bh$ is faithful. Let $\hat{T} \mathpunct : \A \rightarrow \bh \otimes \bc{2}$ be a completely positive extension for the cp maps $\sigma_{} \circ T_{i} \mathpunct : \A \rightarrow \bh$. Since $\B$ is injective, we can find a completely positive conditional expectation $P \mathpunct : \bh \rightarrow \sigma(\B)$ and then set $\hat{T}' := (\sigma^{-1}_{} \circ P_{} \, \otimes \id_{2}) \circ \hat{T}$. This defines a completely positive extension for the cp maps $T_{i}$, and from the definition of the Bures distance we then immediately have the estimate
\begin{equation}
    \label{eq:injective02}
        \begin{split}
            \beta^{2}(T_{1}, T_{2}) & \leq \norm{ T_{1}(\1_{\A}) + T_{2}(\1_{\A}) - \hat{T}_{12}' (\1_{\A}) - \hat{T}_{21}' (\1_{\A})}\\
            & =\norm{\sigma^{-1} \circ P \circ \big ( \sigma_{} \circ T_{1}(\1_{\A}) + \sigma_{} \circ T_{2}(\1_{\A}) -
            \hat{T}_{12} (\1_{\A}) - \hat{T}_{21} (\1_{\A}) \big )} \\
            & \leq \norm{\sigma_{} \circ T_{1}(\1_{\A}) + \sigma_{} \circ T_{2}(\1_{\A}) -
            \hat{T}_{12} (\1_{\A}) - \hat{T}_{21} (\1_{\A})} \, ,
        \end{split}
\end{equation}
where in the last step we have used that both $\sigma^{-1}$ and $P^{}$ are completely positive with norm $\leq 1$. Since Eq.~(\ref{eq:injective02}) holds for all extensions of $\sigma_{} \circ T_{i}$, we conclude that
\begin{equation}
    \label{eq:injective03}
        \beta(T_{1}, T_{2}) \leq \beta(\sigma_{} \circ T_{1}, \sigma_{} \circ T_{2})
\end{equation}
for any faithful representation $\sigma$, as suggested. $\blacksquare$

With the help of Prop.~\ref{propo:injective}, the proof of Th.~\ref{theo:cont02} can now be obtained directly from Th.~\ref{theo:cont01}.

{\bf Proof of Th.~\ref{theo:cont02}:} Since the cb-norm is invariant under faithful representations, Eq.~(\ref{eq:main05}) immediately follows by choosing the representation $\sigma$ to be faithful in Eq.~(\ref{eq:injective01}) and applying the corresponding bound from Th.~\ref{theo:cont01}. If in addition the range algebra $\B$ is injective, Eq.~(\ref{eq:main06}) follows in the same way from Prop.~\ref{propo:injective} and Th.~\ref{theo:cont01}. $\blacksquare$


\section{Appendix A: Properties of the Bures Distance}
    \label{sec:distance}

In this Section we will show that the Bures distance $\beta(T_{1},
T_{2})$ defined in Eq.~(\ref{eq:main02}) indeed has all the
properties of a distance measure.

\begin{propo}
    \label{propo:distance}
    {\bf (Bures Distance)}\\
    The functional $(T_{1}, T_{2}) \mapsto \beta(T_{1},T_{2})$ is
    a metric on the set of cp maps $T_{i} \mathpunct : \A \rightarrow \bh$.
\end{propo}
{\bf Proof:} Positivity and symmetry are immediate from the
definition of $\beta(T_{1},T_{2})$. Obviously, $\beta(T_{1},T_{1})
= 0$. Conversely, Prop.~\ref{propo:lower} shows that
$\beta(T_{1},T_{2}) = 0$ entails $\cb{T_{1}-T_{2}} = 0$, and hence
$T_{1} = T_{2}$. Thus, it only remains to establish the triangle
inequality, $\beta(T_{1}, T_{3}) \leq \beta(T_{1}, T_{2}) +
\beta(T_{2}, T_{3})$ for all triples of cp maps $T_{i}$. To this
end, let $(\pi_{},V_{i},\kk)$ be dilations for the cp maps $T_{1}$
and $T_{2}$ with a common representation $\pi$. Further assume
that $(\check{\pi}_{},\check{V}_{j},\check{\kk})$ are dilations
for the pair $T_{2},T_{3}$ with a common representation
$\check{\pi}$. As before,
$(\hat{\pi}_{i},\hat{V}_{i},\hat{\kk}_{i})$ will denote the
corresponding minimal dilations, with intertwiners $U_{i}
\mathpunct : \hat{\kk}_{i} \rightarrow \kk_{}$ for $i\in \{1,2\}$
and $\check{U}_{j} \mathpunct : \hat{\kk}_{j} \rightarrow
\check{\kk}_{}$ for $j\in \{2,3\}$. We now set
\begin{align}
    \label{eq:distance01}
            \tilde{V}_{1} \ket{\psi} & := \sqrt{\1_{\hat{\kk}_{1}} -
            U_{1}^{*} U_{2}^{} U_{2}^{*} U_{1}^{}} \, \hat{V}_{1}
            \ket{\psi} \; \oplus \; U_{2}^{*} V_{1} \ket{\psi} \; \oplus \; 0
            \, , \\
    \label{eq:distance02}
            \tilde{V}_{2} \ket{\psi} & := 0 \; \oplus \;
            \hat{V}_{2} \ket{\psi} \; \oplus \; 0 \, , \\
    \label{eq:distance03}
            \tilde{V}_{3} \ket{\psi} & := 0 \; \oplus \;
            \check{U}_{2}^{*} \check{V}_{3}^{} \ket{\psi} \;
            \oplus \; \sqrt{\1_{\hat{\kk}_{3}} -
            \check{U}_{3}^{*} \check{U}_{2}^{} \check{U}_{2}^{*}
            \check{U}_{3}^{}} \, \hat{V}_{3}^{} \ket{\psi} \, .
\end{align}
Obviously, $\tilde{V}_{2} \in S(T_{2}, \hat{\pi}_{1} \oplus
\hat{\pi}_{2} \oplus \hat{\pi}_{3})$. Making use of the intertwining relations
Eq.~(\ref{eq:intro03}), we also have
\begin{equation}
    \label{eq:distance03a}
        \begin{split}
            \tilde{V}_{1}^{*} \, & \big ( \hat{\pi}_{1}(a) \oplus
            \hat{\pi}_{2}(a) \oplus
            \hat{\pi}_{3}(a) \big ) \, \tilde{V}_{1}^{}\\
            & = \hat{V}_{1}^{*}
            \sqrt{\1_{\hat{\kk}_{1}} - U_{1}^{*} U_{2}^{}
            U_{2}^{*} U_{1}} \, \hat{\pi}_{1}(a) \,
            \sqrt{\1_{\hat{\kk}_{1}} - U_{1}^{*} U_{2}^{}
            U_{2}^{*} U_{1}} \hat{V}_{1}^{} \, + \, V_{1}^{*} \, U_{2}^{} \, \hat{\pi}_{2}(a) \,
            U_{2}^{*} \, V_{1}^{}\\
            & = V_{1}^{*} \, U_{1} \,
            \hat{\pi}_{1}(a) \, \big ( \1_{\hat{\kk}_{1}} - U_{1}^{*} U_{2}^{}
            U_{2}^{*} U_{1} \big ) \, U_{1}^{*} \, V_{1}^{} \, + \, V_{1}^{*} \, \pi_{}(a) \, U_{2}^{} \, U_{2}^{*} \,
            V_{1}^{} \\
            & = V_{1}^{*} \, U_{1}^{} \, U_{1}^{*} \,
            \pi(a) \, \big ( \1_{\kk} - U_{2}^{} \, U_{2}^{*} \big
            ) \, U_{1}^{} \, U_{1}^{*} \, V_{1}^{} \, + \, V_{1}^{*} \, \pi_{}(a) \, U_{2}^{} \, U_{2}^{*} \,
            V_{1}^{} \\
            & = V_{1}^{*} \,
            \pi(a) \, \big ( \1_{\kk} - U_{2}^{} \, U_{2}^{*} \big
            ) \, V_{1}^{} \, + \, V_{1}^{*} \, \pi_{}(a) \, U_{2}^{} \, U_{2}^{*} \,
            V_{1}^{} = T_{1} (a)
        \end{split}
\end{equation}
for all $a \in \A$, and thus $\tilde{V}_{1} \in S(T_{1},
\hat{\pi}_{1} \oplus \hat{\pi}_{2} \oplus \hat{\pi}_{3}).$ An analogous calculation shows that $\tilde{V}_{3} \in S(T_{3},
\hat{\pi}_{1} \oplus \hat{\pi}_{2} \oplus \hat{\pi}_{3}).$ Hence, $\hat{\pi}_{1} \oplus \hat{\pi}_{2} \oplus \hat{\pi}_{3}$ is a
common representation for the completely positive maps $T_{1}, T_{2}$, and $T_{3}$ with corresponding dilations $(\tilde{V}_{i}, \hat{\pi}_{1} \oplus
\hat{\pi}_{2} \oplus \hat{\pi}_{3}, \hat{\kk}_{1} \oplus
\hat{\kk}_{2} \oplus \hat{\kk}_{3})$. Moreover, we see from
Eq.~(\ref{eq:distance02}) that $\tilde{V}_{2}$ only depends on
the minimal dilations. In addition, we have
\begin{equation}
    \label{eq:distance04}
        \tilde{V}_{2}^{*} \, \tilde{V}_{1}^{} = V_{2}^{*} \,
        V_{1}^{} \qquad {\rm and} \qquad
        \tilde{V}_{2}^{*} \, \tilde{V}_{3}^{} = \check{V}_{2}^{*} \,
        \check{V}_{3}^{} \, .
\end{equation}
Now assume that $(\pi_{},V_{i},\kk)$ and
$(\check{\pi}_{},\check{V}_{j},\check{\kk})$ are chosen as in
Prop.~\ref{propo:upper} such that
\begin{align}
    \label{eq:distance05}
        \norm{V_{1} - V_{2}} & = \beta_{\pi}(T_{1}, T_{2}) = \beta_{}(T_{1},
        T_{2}) \, ,\\
    \label{eq:distance06}
        \norm{\check{V}_{2} - \check{V}_{3}} & = \beta_{\check{\pi}}
        (T_{2}, T_{3}) = \beta_{}(T_{2}, T_{3}) \, .
\end{align}
Hence, Eq.~(\ref{eq:distance04}) and the triangle inequality for
the operator norm imply that
\begin{equation}
    \label{eq:distance07}
        \begin{split}
            \beta (T_{1},T_{3}) & \leq \norm{\tilde{V}_{1} -
            \tilde{V}_{3}}\\
            & \leq \norm{\tilde{V}_{1} - \tilde{V}_{2}} +
            \norm{\tilde{V}_{2} - \tilde{V}_{3}}\\
            & = \norm{V_{1} - V_{2}} + \norm{\check{V}_{2} -
            \check{V}_{3}}\\
            & = \beta (T_{1}, T_{2}) + \beta ( T_{2}, T_{3}) \, ,
        \end{split}
\end{equation}
concluding the proof. $\blacksquare$


\section{Appendix B: Bures's Upper Bound for Positive Functionals}
    \label{sec:functional}

The proof of the upper bound $\beta(T_{1}, T_{2}) \leq
\sqrt{\norm{T_{1} - T_{2}}}$ for cp maps $T_{i}$ that we
present in Sec.~\ref{sec:upper} relies on the corresponding result
for positive functionals. In his original paper \cite{Bur69} Bures
assumed (normalized) states on von Neumann algebras. The
generalization to arbitrary bounded positive functionals on
$C^{*}$-algebras is straightforward. We nevertheless include it
here for completeness and reference.

\begin{propo}
    \label{propo:bures}
        {\bf (Bures's Bound for Positive Functionals)}\\
        Let $\A$ be a $C^{*}$-algebra, and let $\omega_{0},
        \omega_{1} \in \A^{*}$ be positive functionals. We then have
        \begin{equation}
            \label{eq:functional01}
                \beta(\omega_{0}, \omega_{1}) \leq
                \sqrt{\norm{\omega_{0} - \omega_{1}}} \; .
        \end{equation}
\end{propo}
The following lemma will establish Prop.~\ref{propo:bures} under
an additional dominance condition. This extra condition will then be removed
with the help of Lemma~\ref{lemma:mixture}, which proves the
continuity of the Bures distance with respect to convex mixtures.

\begin{lemma}
    \label{lemma:functional01}
        Let $\A$ be a $C^{*}$-algebra, and let $\omega_{0},
        \omega_{1} \in \A^{*}$ be two positive functionals
        such that $n \, \omega_{0} \geq \omega_{1}$ for some $n \in \N$.
        Then Eq.~(\ref{eq:functional01}) holds.
\end{lemma}
{\bf Proof of Lemma~\ref{lemma:functional01}:} We choose a common representation $\pi \mathpunct : \A \rightarrow \bkk$ such that the fibers $S(\omega_{i},\pi)$ are non-empty, $i \in \{0,1\}$. The functionals $\omega_{i}$ admit unique normal extensions to the von Neumann algebra $\A^{\pi}:=\pi(\A)''$, which we denote by $\omega_{i}^{\pi}$. The dominance condition transfers, hence $n \, \omega^{\pi}_{0} - \omega^{\pi}_{1} \geq 0$ for some $n \in \N$. Sakai's Radon-Nikodym theorem \cite{Sak65,Sak71} then allows us to find a positive
operator $h \in \A^{\pi}$ such that $\omega^{\pi}_{1}(a) = \omega^{\pi}_{0}(h a h)$
for all $a \in \A$. For $\ket{\psi} \in S(\omega_{0},\pi_{})$ we
have $h \ket{\psi} \in S(\omega_{1}, \pi_{})$, and hence
\begin{equation}
    \label{eq:functional02}
        \beta^{2}(\omega_{0}, \omega_{1}) \leq \norm{(\1 - h)
        \ket{\psi}}^{2} = \omega^{\pi}_{0} \big ( (\1 - h)^2 \big ) \, .
\end{equation}
Let $h = \int \lambda \, p(d\!\lambda)$ denote the spectral
decomposition of $h$, and set $p := \int_{\lambda=0}^{1}
p(d\!\lambda)$. We then find
\begin{align}
    \label{eq:functional03}
        \omega^{\pi}_{0} \big ( (\1 - h)^2 \, p \big ) & \leq \omega^{\pi}_{0} \big ( (\1 - h^{2}) \, p \big ) \quad {\rm and} \\
    \label{eq:functional04}
        \omega^{\pi}_{0} \big ( (\1 - h)^{2} \, (\1-p) \big ) & \leq
        \omega^{\pi}_{0} \big ( (\1 - h^{2}) \, (p-\1) \big ) \, .
\end{align}
Adding Eqs.~(\ref{eq:functional03}) and (\ref{eq:functional04}),
we see from Eq.~(\ref{eq:functional02}) that
\begin{equation}
    \label{eq:functional05}
        \begin{split}
            \beta^{2} (\omega_{0}, \omega_{1}) & \leq \omega^{\pi}_{0} \big (
            (\1 - h^{2}) \, p \big ) + \omega^{\pi}_{0} \big ( (\1 - h^{2}) \, (p-\1) \big
            ) \\
            & = \omega^{\pi}_{0} \big ( (\1-h^{2}) \, ( 2 \, p - \1)
            \big)\\
            & = (\omega^{\pi}_{0} - \omega^{\pi}_{1}) ( 2 \, p - \1)\\
            & \leq \norm{\omega^{\pi}_{0} - \omega^{\pi}_{1}}\\
            & = \norm{\omega_{0} - \omega_{1}} \, ,
        \end{split}
\end{equation}
where we have used that $2 \, p - \1 = p - (\1 -
p)$ is a reflection, and hence $\norm{2 \, p - \1} = 1$.
$\blacksquare$

\begin{lemma}
    \label{lemma:mixture}
        Let $\A$ be a $C^{*}$-algebra, and let $\omega_{0},
        \omega_{1} \in \A^{*}$ be two positive functionals. Then
        the inequality
        \begin{equation}
            \label{eq:functional06}
                \abs{ \beta \big (\omega_{0}, \omega_{1} \big) - \beta
                \big ( \, (1-s) \,
                \omega_{0} + s \, \omega_{1}, \omega_{1} \big )} \leq \sqrt{s} \, \big (
                \sqrt{\norm{\omega_{0}}} +
                \sqrt{\norm{\omega_{1}}} \big )
        \end{equation}
        holds for all $s \in [0,1]$.
\end{lemma}
{\bf Proof of Lemma~\ref{lemma:mixture}:} Again, the proof
proceeds via a direct sum construction. For $\ket{\psi_{i}} \in
S(\omega_{i}, \pi_{})$ we have $\ket{\psi_{0}} \oplus 0 \in
S(\omega_{0}, \pi_{} \oplus \pi_{})$ and $\ket{\psi_{0}} \oplus
\ket{\psi_{1}} \in S(\omega_{0} + \omega_{1}, \pi_{} \oplus
\pi_{})$, and thus
\begin{equation}
    \label{eq:functional07}
        \beta (\omega_{0}, \omega_{0} + \omega_{1}) \leq
        \norm{\,\ket{\psi_{0}} \oplus 0 - \ket{\psi_{0}} \oplus
        \ket{\psi_{1}}} = \bk{\psi_{1}} =
        \sqrt{\norm{\omega_{1}}} \, .
\end{equation}
We know from Prop.~\ref{propo:distance} that the Bures distance is
indeed a metric, and hence we can use the triangle inequality and
then Eq.~(\ref{eq:functional07}) to conclude that
\begin{equation}
    \label{eq:functional08}
        \begin{split}
            | \beta \big (\omega_{0}, \omega_{1} \big) & - \beta
            \big ( \, (1-s) \, \omega_{0} + s \, \omega_{1}, \omega_{1} \big
            )| \leq \beta \big ( \omega_{0}, (1-s) \, \omega_{0} + s \, \omega_{1}
            \big )\\
            & \leq \beta \big ( \omega_{0}, (1-s) \omega_{0} \big
            ) + \beta \big ( (1-s) \omega_{0}, (1-s) \, \omega_{0} + s \, \omega_{1}
            \big )\\
            & \leq \sqrt{s} \, \big ( \sqrt{\norm{\omega_{0}}} +
            \sqrt{\norm{\omega_{1}}} \big ) \, ,
        \end{split}
\end{equation}
just as suggested. $\blacksquare$

We now have all the necessary tools at hand for the

{\bf Proof of Prop.~\ref{propo:bures}:} Given a parameter $s\in (0,1]$, we define the convex mixture $\omega_{s} := (1-s) \, \omega_{0} + s \, \omega_{1}$. Choosing a
positive integer $n > s^{-1}$, we have $n \, \omega_{s} -
\omega_{1} > 0$, and hence $\beta ( \omega_{s}, \omega_{1} ) \leq
\sqrt{\norm{\omega_{s} - \omega_{1}}}$ follows from
Lemma~\ref{lemma:functional01}. We can then conclude from
Lemma~\ref{lemma:mixture} that the estimate
\begin{equation}
    \label{eq:functional09}
        \begin{split}
            \beta( \omega_{0}, \omega_{1}) & \leq \beta (
            \omega_{s}, \omega_{1} ) + \sqrt{s} \, \big (
            \sqrt{\norm{\omega_{0}}} + \sqrt{\norm{\omega_{1}}}
            \big ) \\
            & \leq \sqrt{\norm{\omega_{s} - \omega_{1}}} + \sqrt{s} \, \big (
            \sqrt{\norm{\omega_{0}}} + \sqrt{\norm{\omega_{1}}}
            \big ) \\
            & = \sqrt{1-s} \, \sqrt{\norm{\omega_{0}
            - \omega_{1}}} + \sqrt{s} \, \big (
            \sqrt{\norm{\omega_{0}}} + \sqrt{\norm{\omega_{1}}}
            \big )
        \end{split}
\end{equation}
holds for all $s \in (0,1]$. The limit $s \rightarrow 0$ yields
the desired result. $\blacksquare$


\section*{Acknowledgments}

We would like to thank Mauro D'Ariano and Vern Paulsen for fruitful and stimulating discussions.

DS acknowledges financial support from Consorzio Nazionale
Interuniversitario per le Scienze della Materia (CNISM). DK is grateful for generous support from Deutscher Akademischer Austauschdienst (DAAD).



\end{document}